\documentclass[aps,prb,twocolumn,superscriptaddress,showpacs,amsmath,floatfix,citeautoscript]{revtex4}

\usepackage{graphicx}  % needed for figures
\usepackage{subfigure}
\usepackage{dcolumn}   % needed for some tables
\usepackage{bm}        % for math
\usepackage{amssymb}   % for math
\usepackage{mathtools}
\usepackage{epstopdf}
\usepackage{color}
\usepackage{ulem}

\begin{document}

\title{Solving frustrated quantum many-particle models with convolutional neural networks}

\author{Xiao Liang}
\affiliation{Laboratory of Quantum Information, University of Science and Technology of China, Hefei, 230026, China}
\affiliation{Synergetic Innovation Center of Quantum Information and Quantum Physics, University of Science and Technology of China, Hefei, 230026, China}
\author{Wen-Yuan Liu}
\affiliation{Laboratory of Quantum Information, University of Science and Technology of China, Hefei, 230026, China}
\affiliation{Synergetic Innovation Center of Quantum Information and Quantum Physics, University of Science and Technology of China, Hefei, 230026, China}
\affiliation{Department of Physics, The Chinese University of Hong-Kong, Shatin, New Territories, Hong Kong}
\author{Pei-Ze Lin}
\affiliation{Laboratory of Quantum Information, University of Science and Technology of China, Hefei, 230026, China}
\affiliation{Synergetic Innovation Center of Quantum Information and Quantum Physics, University of Science and Technology of China, Hefei, 230026, China}
\author{Guang-Can Guo}
\affiliation{Laboratory of Quantum Information, University of Science and Technology of China, Hefei, 230026, China}
\affiliation{Synergetic Innovation Center of Quantum Information and Quantum Physics, University of Science and Technology of China, Hefei, 230026, China}
\author{Yong-Sheng Zhang}
\email{yshzhang@ustc.edu.cn}
\affiliation{Laboratory of Quantum Information, University of Science and Technology of China, Hefei, 230026, China}
\affiliation{Synergetic Innovation Center of Quantum Information and Quantum Physics, University of Science and Technology of China, Hefei, 230026, China}
\author{Lixin He}
\email{helx@ustc.edu.cn}
\affiliation{Laboratory of Quantum Information, University of Science and Technology of China, Hefei, 230026, China}
\affiliation{Synergetic Innovation Center of Quantum Information and Quantum Physics, University of Science and Technology of China, Hefei, 230026, China}
\date{\today}

\begin{abstract}
Recently, there has been significant progress in solving quantum many-particle problem via machine learning
based on the restricted Boltzmann machine. However, it is still highly challenging to solve frustrated models
via machine learning, which has not been demonstrated so far. In this work, we design
a brand new convolutional neural network (CNN) to solve such quantum many-particle problems.
We demonstrate, for the first time, of solving the highly frustrated spin-1/2 J$_1$-J$_2$ antiferromagnetic Heisenberg model on square lattices via CNN.
The energy per site achieved by the CNN is even better than previous string-bond-state calculations.
Our work therefore opens up a new routine to solve challenging frustrated quantum many-particle problems using machine learning.
\end{abstract}
\maketitle

\section{Introduction}

The successes of machine learning in image recognition, \cite{Imagenet} playing Go, \cite{GO_1, GO_2} etc., stimulate interests in using machine learning techniques to solve physics problems, such as designing optical experiments, \cite{melnikov} processing signals in finding gravity waves, \cite{gravity_wave_sim,gravity_wave_prl} quantum teleportation photon spot recognition, \cite{teleportation} quantum phase distinguishing, \cite{phase_1,phase_2,phase_3,phase_4,phase_5,phase_6} inferring Hamiltonian solutions, \cite{kmills_1,kmills_2} classifying quantum states based on topological invariants, \cite{topo_invariant} classifying spin configurations into phases, \cite{CNN_phase} solving ground states of Bose-Hubbard model, \cite{CNN_bose_hubbard} and so on.

Recently machine learning has been applied to study quantum many-particle problems. \cite{Carleo, DDL} This posed serious challenges, because it requires much higher precision than traditional machine learning problems. It has been demonstrated that by unsupervised learning, the restricted Boltzmann machine (RBM) can solve the ground states of transverse-field Ising models and anti-ferromagnetic Heisenberg models in rather high precision.\cite{Carleo}
The ground states of those models obey
the Marshall-Peierls sign rule (MPSR) \cite{mpsr} where the wave functions  can be represented by positive numbers, suitable for RBM with real parameters.

The representation ability of RBM for some advanced states has been investigated, and often connects to another state-of-the-art method for many-particle problems, namely the tensor network state (TNS) method\cite{TNS}. The equivalence between RBM and TNS has been investigated in Refs.\cite{DDL,PRX,PRB}. It has been suggested that RBM can represent the quantum states beyond area law. \cite{DDL} Furthermore, the extension of RBM, namely deep Boltzmann machine (DBM), can be transformed into tensor networks, and DBM can be a general representation of quantum many-particle states. \cite{Gaoxun}

In the community of machine learning, it is believed that
convolutional neural networks (CNN) have shown more efficiency compared
to the neural networks within dense connections. \cite{CNN_advance} Recent works try to bridge the state-of-the-art convolutional neural network to the tensor network states.~\cite{CNN_TNS} It has been proposed that the information reuse introduced by overlapping convolution filters can enhance the state representation ability. The convolutional neural network has shown effectiveness in finding the critical points in phase transitions of Ising models, \cite{phase_nat_phys} which infers the network's valid state representation ability.

However, so far solving frustrated quantum many-particle models on large two-dimensional lattices is
still challenging for neural networks. The ability to solve the frustrated model is a milestone for neural networks, because
the non-frustrated models can be efficiently solved to extremely high accuracy by quantum Monte Carlo (QMC) methods.
In this work, we built a CNN to solve the ground states of two-dimensional frustrated quantum spin models, more specifically the antiferromagnetic spin-1/2 J$_1$-J$_2$ Heisenberg model on square lattices. This model is highly frustrated, and considered as one of the most interesting and challenging spin models.\cite{SBS,J1J2_1} The network is built using the elements from state-of-the-art neural networks. By design, the network associates
the many-body wave functions with high-order spin correlation functions.
We name such CNN as the convolutional quantum neural state (CQNS). We solve the ground states at $J_2=0$ and $J_2/J_1=0.5$ and it is shown that CQNS can achieve the ground energies lower than those obtained by the previous string bond state (SBS) calculations.~\cite{SBS,Replica_exchange} We further compare the spin correlation functions, focusing on the convolution filters' size effects. We show that if the size of the filter is too small, CQNS can not capture the correct long-range spin correlations, which leads to poor energy precisions. However, for systems with the correlation length shorter than the system size, it is efficient to use relative smaller convolution filters.
The CQNS is a general method  which can be easily applied to study other quantum many-particle problems.

\section{Network Structure}
\label{sec:level2}

\begin{figure}
\includegraphics[width=0.45\textwidth]{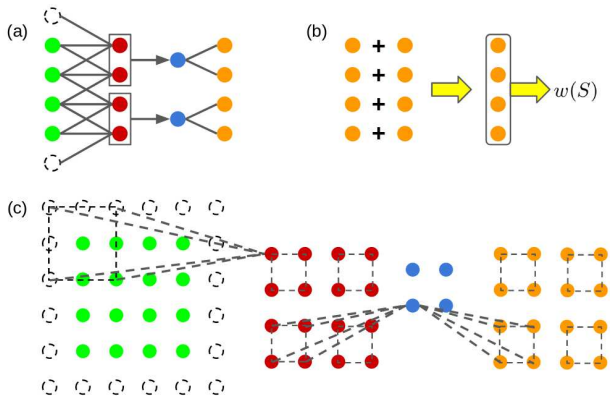}
\caption{ (Color online) (a) The structure of a CQNS for a periodic spin chain with length $L=4$,
and the convolution filter length $K=3$. Because of padding, the convolution filter scans on the chain length of six.
The green neurons are the input states whereas the red neurons are the outputs after convolution. Max-pooling is performed by two neurons with the stride number of two, where the outputs are the blue neurons. The orange neurons are the outputs after transposed convolution. (b)The final chain of neurons is the element-wise summation of the output chains from all convolution filters.
%, i.e., a neuron in the final chain is the summation of the neurons on the same location from all output chains, while each convolution filter generates one output chain.
(c) An example of a two-dimensional CQNS on a 4$\times$4 square lattice with periodic boundary condition and the size of
the convolution filter is $K=3$. The max-pooling and transposed convolution are also performed on
the two-dimensional lattice. }
\label{fig1}
\end{figure}

To illustrate the structure of CQNS, we begin with an example of a one-dimensional spin chain. The spin chain has finite length $L=4$ with periodic boundary conditions (PBC). The size of the convolution filter is taken to be $K=3$. The network structure is shown in Fig.~\ref{fig1}(a). The input spins are labeled as the green neurons, which are presented
in the Pauli matrix $\sigma_z$ base. To keep the output dimension unchanged after convolution, we pad additional spin sites on the head and the tail of the spin chain according to PBC. The number of sites padded on each side of the spin chain is $(K-1)/2$. For $K=3$, one additional spin site is padded on both the head and the tail of the spin chain.
The reason to keep the output dimension unchanged after convolution is to use the advantage of max-pooling. In image recognition tasks, max-pooling reduces the output dimensions and introduces position invariance \cite{pooling}, thus it is possible to use a small number of convolution filters to grasp all possible spin configurations in the receptive field, especially when the filter size is large. The output neurons after max-pooling are denoted by the blue neurons. After max-pooling, the dimension is restored to the dimension of the spin lattice by a transposed convolution.

A quantum state of a spin lattice can be represented in the form of $|\Psi\rangle=\sum_{\bf S} w({\bf S})|{\bf S}\rangle$, where $|{\bf S}\rangle=|s_1,s_2,\cdots,s_N\rangle$ is a spin configuration, and $w({\bf S})$ is its coefficient. For a given spin configuration $|{\bf S}\rangle$, CQNS outputs the wave function coefficient $w({\bf S})$, which is the products of all the final output neurons. Figure~\ref{fig1}(a) depicts the network structure of using one convolution filter, which generates one chain of output neurons. When $M$ convolution filters are used, there will be $M$ chains of output neurons.
To correlate different filters, we sum all the output neurons on the same site generated from all filters, as Fig.~\ref{fig1}(b) denotes. The wave function coefficient $w({\bf S})$ is obtained by taking the products of all the neurons in the final chain.

One can easily extend the above one-dimensional neural network to higher dimensions. Figure~\ref{fig1}(c) depicts a two-dimensional CQNS with a $K\times K$ convolution filter, with $K=3$. After convolution layer, a max-pooling is performed on adjacent four neurons followed by a transposed convolution restores the dimensions. The wave function coefficient $w({\bf S})$ is the product of all the neurons on the final output plane.

To reveal how the CQNS works, we use the $L$=4 spin chain as an example. Considering a convolution filter with length $K=3$ and one additional spin padding in both the head and the tail of the spin chain,
the convolution operation can be written in the matrix form:
\begin{equation}
\begin{bmatrix}
p_1\\
p_2\\
p_3\\
p_4
\end{bmatrix}=
\textbf{b}+
\begin{bmatrix}
w_1 &w_2 &w_3 &0 &0 &0\\
0 &w_1 &w_2 &w_3 &0 &0\\
0 &0 &w_1 &w_2 &w_3 &0\\
0 &0 &0 &w_1 &w_2 &w_3
\end{bmatrix}
\begin{bmatrix}
s_4\\
s_1\\
\vdots \\
s_1
\end{bmatrix},
\label{2}
\end{equation}
where $\textbf{S}$=$[s_4, s_1, s_2, s_3, s_4, s_1]^\mathsf{T}$ is the input spin configuration and $\textbf{b}$ is the bias vector of the convolution filter. In our cases, the input number of each site is $s_i$=$\pm 1$. The output neurons after convolution are the input neurons of the max-pooling, i.e.,
\begin{equation}
\begin{bmatrix}
g_1\\
g_2
\end{bmatrix}=
\begin{bmatrix}
\text{max}(p_1,p_2)\\
\text{max}(p_3,p_4)
\end{bmatrix} \, .
\label{3}
\end{equation}
We then perform transposed convolution, 
\begin{equation}
\begin{bmatrix}
h_1\\
h_2\\
h_3\\
h_4
\end{bmatrix}=
\begin{bmatrix}
d_1 &0\\
d_2 &0\\
0 &d_1\\
0 &d_2
\end{bmatrix}
\begin{bmatrix}
g_1\\
g_2
\end{bmatrix}.
\label{4}
\end{equation}
to restore the original size of the lattice.

Equations~(\ref{2}-~\ref{4}) describe the transformations between the input spins and the output neurons when using one convolution filter. Using $M$ convolution filters lead to $M$ output chains, as shown in Fig.~\ref{fig1}(b).
The final output chain is the direct summation of the output chains from all convolution filters, i.e.,
\begin{equation}
\begin{bmatrix}
h_1\\
h_2\\
h_3\\
h_4
\end{bmatrix}=
%_{\text{final}}=
\sum_{m=1}^{M}
\begin{bmatrix}
d_1^{(m)}g_1^{(m)}\\
d_2^{(m)}g_1^{(m)}\\
d_1^{(m)}g_2^{(m)}\\
d_2^{(m)}g_2^{(m)}
\end{bmatrix}.
\label{5}
\end{equation}
The final wave function coefficient is the product of all neurons in the final output chain, i.e., $w({\bf S})=h_1h_2h_3h_4$. From the above equations, it is easy to see that for a given spin configuration ${\bf S}$
of a $N$ sites system, $h_i$  is linear combination of $s_k$ plus a constant, i.e., $h_i=A_{i,1}s_1+A_{i,2}s_2+\cdots+A_{i,N}s_N+c_i$,
where $A_{i,k}$ is the coefficient of spin $s_k$ in $h_i$, and $c_i$ is a number.
However, due to the max-pooling, the coefficients $A_{i,k}$ are spin configuration dependent.
Therefore we have,
\begin{equation}\label{eq:ws}
 w({\bf S})=\sum_{n_1,\cdots,n_N} g(n_1,\cdots,n_N;s_1,\cdots,s_N)s_1^{n_1}\cdots s_N^{n_N},
\end{equation}
where $n_1+n_2+\cdots +n_N \leq  N$, and $g(n_1,\cdots,n_N;s_1,\cdots,s_N)$ are given by CQNS.
We note that $g$ functions are also spin configuration dependent, because of the max-pooling.
Equation~\ref{eq:ws} associates the many-body wave functions with the high order
spin correlation functions. This is crucial for the success of CQNS that are
very different from traditional CNN, in which
the nonlinearity of the neural networks is usually introduced by the
activation functions.

\begin{figure}[tb]
\includegraphics[width=0.45\textwidth]{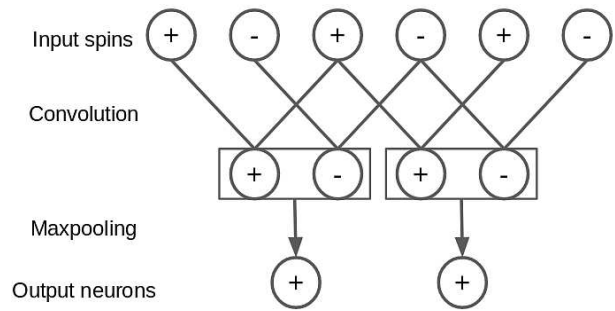}
\caption{ Diagram of CQNS with only one convolution filter. The kernels of the filter are $[0.5, 0, 0.5]$. When the input spin configuration is a N\'eel state, the output neurons after convolution are also positive and negative distributed alternatively. After the max-pooling, all the output neurons are positive, which leads to a non-zero wave function coefficient. For all other input spin configurations, the coefficients are zero. }
\label{fig2}
\end{figure}

It is easy to show that CQNS can exactly represent the classical N\'eel state, which has alternating spin up and spin down
in the lattice, and is the ground state of a classical antiferromagnetic Heisenberg model, using a single convolution filter. The filter kernel has the form $[0.5, 0, 0.5]$, and biases are taken to be zero.
Figure~\ref{fig2} depicts the intermediate neurons after convolution and max-pooling when the input state is a N\'eel state.
The output neurons after max-pooling are all positive, thus the wave function coefficient $w({\bf S})>$0.
It is easy to verify that at least one of the output neurons after max-pooling is zero if the input state is
not the N\'eel state, and therefore $w({\bf S})$=0.
For a quantum antiferromagnetic Heisenberg model more filters are needed to represent
the ground state. As an example, for the spin chain of $L=4$, it is easy to verify that the ground state can be exactly represented by the CQNS with four filters, which are $[-0.5,0.5,0.5]$, $[0.5,0.5,-0.5]$, $[0.5,-0.5,0.5]$ and $[0.5,0,0.5]$ and biases for the filters are -1.5, -1.5, -0.5 and 0 respectively.

\section{Numerical investigations on J$_1$-J$_2$ model}
\label{sec:level4}

We benchmark the CQNS on the two-dimensional spin-1/2 J$_1$-J$_2$ Heisenberg model,
\begin{equation}
H=\sum_{\langle i,j\rangle} J_1\textbf{s}_i\cdot\textbf{s}_j+J_2\sum_{\langle\langle i,j\rangle\rangle}\textbf{s}_i\cdot\textbf{s}_j,
\label{7}
\end{equation}
where $\langle i,j\rangle$ denotes the nearest neighbour sites and $\langle\langle i,j\rangle\rangle$ denotes the next nearest neighbour sites. We assume both $J_1$, $J_2$$>$0, and take $J_1$=1 in all calculations in this work.
The J$_1$-J$_2$ model is one of the most interesting and challenging quantum spin models due to the strong frustrated interactions, and the ground state near
$J_2/J_1=0.5$ is still under intensive debate.~\cite{J1J2_2,J1J2_3,J1J2_4,J1J2_5}
In this work, we only focus on the ground state energies of the model.

%has became a promising candidate model whose ground state may be a spin liquid state near

The total energy of the system can be calculated as:
\begin{equation}
E=\frac{\langle\Psi|H|\Psi\rangle}{\langle\Psi|\Psi\rangle}=\frac{1}{\sum_\textbf{S}w_\textbf{S}^2}\sum_\textbf{S} w_\textbf{S}^2\sum_{\textbf{S}'} \frac{w_{\textbf{S}'}}{w_\textbf{S}}H_{\textbf{S}'\textbf{S}},
\label{8}
\end{equation}
where $H_{\textbf{S}'\textbf{S}}=\langle \textbf{S} |H | \textbf{S}' \rangle$.
Our goal is to minimize the total energy, which can be achieved by the stochastic
gradient method\cite{Gradient},
where the energy $E$ and the gradients $G$ are calculated via Monte Carlo sampling over spin configurations,
\begin{equation}
\begin{split}
&E=\langle E_\textbf{S}\rangle,\\
&G=\langle O_\textbf{S}E_\textbf{S}\rangle-\langle E_\textbf{S}\rangle\langle O_\textbf{S}\rangle,
\end{split}
\label{9}
\end{equation}
where $E_S=\sum_{\textbf{S}'} \frac{w({\textbf{S}'})}{w(\textbf{S})} H_{\textbf{S}'\textbf{S}}$ and $O_\textbf{S}=\frac{1}{w(\textbf{S})}\frac{\partial w(\textbf{S})}{\partial a_i}$.
Here $a_i$ are all the parameters in the CQNS.
\begin{figure*}[tb]
\includegraphics[width=0.75\textwidth]{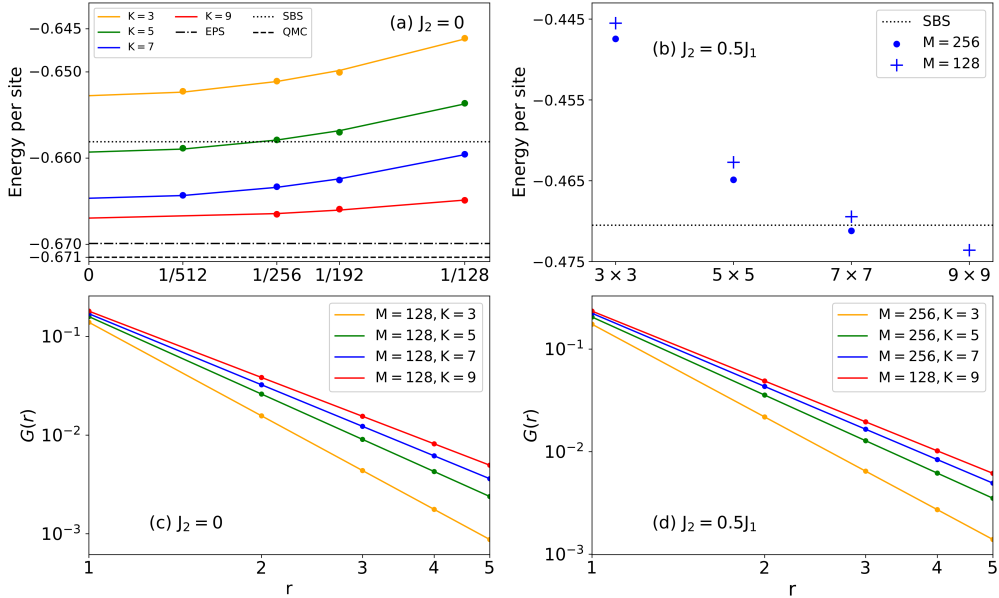}
\caption{ (Color online) (a) The ground state energies calculated by CQNS at $J_2=0$ as functions of the
number of filters $M$ and size of filters $K$. The reference ground energies obtained by the string-bond-state (SBS),
entangled-plaquette-state (EPS) and  QMC are also shown in the dotted line (-0.6581), the dash-dotted line (-0.6699) and the dashed line (-0.6715), respectively.
%For $K=9$, the energy per site from 128 filters to 256 filters is -0.6649, -0.6659 and -0.6665. The solid circles are the %actual values obtained by CQNS and the solid lines denote the fitted function based on the solid circles.
(b) The ground state energies calculated by CQNS at $J_2=0.5 J_1$ as functions of $M$ and $K$. The ground state energy obtained by SBS is shown
as the dotted line (-0.4705). The fitted spin correlation function $G_r$ are shown in (c) for $J_2$=0, and (d) for $J_2$=0.5$J_1$ for different number of filters $M$ and size of filters $K$.}
\label{fig4}
\end{figure*}

However, the neural networks are highly nonlinear, therefore the system may be easily trapped in
some local minima via the stochastic gradient descending method.
To overcome this difficulty, we optimize CQNS via a recently developed replica-exchange molecular dynamics method, which has been successfully applied to optimize the SBS~\cite{Replica_exchange}, one type of TNS. We map the quantum many-particle problem to a classical mechanical problem, in which we treat the parameters of CQNS as the generalized coordinates of the system. We optimize the energy of the virtual system using a replica exchange molecular dynamics method. By exchanging the system configurations among higher and lower temperatures, it can explore large phase space and therefore effectively avoid being stuck in the local minima \cite{Caokun_1,Caokun_2,lorenzo}. Details of the method are presented in Ref. \cite{Replica_exchange}.

In our simulations, we use 56 temperatures. Initially the temperatures distribute exponentially between the highest ($1/\beta_0=10^{-3}$) and lowest ($1/\beta_{55}=10^{-5}$) temperatures. The MD step length is set to $\Delta t=0.01$. For each temperature we start with random weights and biases. During the simulations, we adjust the temperatures after configuration exchange for 10 times, whereas there are 210 MD steps between the exchanges of two configurations. For each MD step, we sample 5000 spin configurations. The energies used for replica exchange are averaged over 200 MD steps.

We study $J_1$-$J_2$ model on a $10\times 10$ two-dimensional square lattice, with periodic boundary conditions. Figure~\ref{fig4} depicts the numerical results for $J_2$=0, and $J_2$=0.5, with different number of filters and filter sizes. The filter sizes are taken to be $K$$\times$$K$. The obtained ground state energies
 are summarized in Table I.

The energies per site
for $J_2$=0 calculated with different number of filters and filter sizes are shown in Fig.~\ref{fig4}(a).
We also show the total energies calculated from other methods as comparisons. The dotted line
denotes the ground state energy per site obtained by the SBS, $E=-0.6581$, \cite{SBS} and the dash-dotted line denotes the result of the entangled plaquette states $E=-0.6699$,\cite{EPS} whereas the dashed line is the result of quantum Monte Carlo $E=-0.6715$.\cite{EPS}  The solid dots are the results obtained by CQNS.
The energy for $K=3$ and $M$=128 filters is -0.6461. The energy can be improved by increasing the number of filters. When using 512 filters, the energy decreases to -0.6522. However for the small filters, further increasing the number of filters does not improve the energy much. On the other hand, increasing filter size can dramatically improve the energy. For example, for filter number $M$=128, the energy is -0.6536 for $K=5$, -0.6596 for $K=7$ and -0.6649 for $K=9$.
As shown in the figure, when $K=9$, the energy per site converges faster than using smaller filters, the energies
for $M$=128, 192 and 256 filters are -0.6649, -0.6659 and -0.6665, respectively.
Based on the numerical results, CQNS with filter size of $K$=7, 9 can exceed SBS. From the numerical results, it is revealed that larger filter number and larger filter size lead to more accurate ground states, however increasing filter size is more efficient than increasing filter number.

At $J_2=0.5$, the model is highly frustrated, where QMC suffers from sign problem, and therefore is much more difficult to calculate. In Fig.~\ref{fig4}(b), we show the calculated energies with different filter size $K$ and number of filters $M$.
The ground state energy per site calculated by SBS is -0.4705, which is shown in dotted line as a comparison.
According to the figure, the energy has limited improvement from $M$=128 to 256 filters. However increasing filter size can significantly improve the energy. For $M$=128, the energy is -0.4455 for $K$=3, -0.4627 for $K$=5 and -0.4736 for $K$=9. For $M$=256, the energy per site is -0.4475 for $K=3$, -0.4649 for $K=5$ and -0.4715 for $K=7$. CQNS can exceeds SBS with filter size $K \ge$7.

\begin{table*}[]
\caption{Comparison of the ground state energies of the two-dimensional $J_1$-$J_2$ Heisenberg model calculated by CQNS. $M$ denotes the convolution filter number and $K$ denotes the side length of the convolution filters.}
\centering
\begin{tabular}{cllllll}
\hline\hline
&\multicolumn{4}{c}{$J_2$=0}                                                                                 & \multicolumn{2}{c}{$J_2$=0.5}                     \\
$M$     & \multicolumn{1}{c}{128} & \multicolumn{1}{c}{192} & \multicolumn{1}{c}{256} & \multicolumn{1}{c}{512} & \multicolumn{1}{c}{128} & \multicolumn{1}{c}{256} \\ \hline
$K$=3 & -0.646095               & -0.650056               & -0.651070               & -0.652247                & -0.445505               & -0.447450                \\
$K$=5 & -0.653637               & -0.657002               & -0.657887               & -0.658851                & -0.462716               & -0.464885                \\
$K$=7 & -0.659552               & -0.662539               & -0.663316               & -0.664313                & -0.469446               & -0.471200                \\
$K$=9 & -0.664892               & -0.665917               & -0.666511               &                          & -0.473591               &                          \\ \hline
\end{tabular}
\label{tab:Energy}
\end{table*}

\begin{table}[tb]
\caption{Comparison of the decay power $\gamma$ of the
spin correlation functions calculated by CQNS. $K$ denotes the side length of the convolution filters.}
\centering
\begin{tabular}{lcccc}
\hline\hline
$K$       & 3 & 5 & 7 & 9 \\
\hline
$J_2$=0   & 3.15                  & 2.61                  & 2.39                  & 2.24                   \\
$J_2$=0.5 & 3.00                  & 2.53                  & 2.37                  & 2.26                   \\ \hline
\end{tabular}
\label{tab:Decay_power}	
\end{table}

We further calculate the spin correlation functions, and the results are shown in Fig.~\ref{fig4}(c) and (d) for
$J_2$=0, and $J_2$=0.5 respectively. The error bars are smaller than the width of the solid lines and therefore not shown.
The spin correlation functions $C(r)$ are calculated as,
\begin{equation}
C(r)= \frac{1}{2L^2} \sum_{i,j}(\langle \textbf{s}_{i+r,j}\textbf{s}_{i,j}\rangle+\langle \textbf{s}_{i,j+r}\textbf{s}_{i,j}\rangle)\, .
\label{10}
\end{equation}
The calculated results show that the correlation functions obey power-law decay. We fit $C(r)$ as $C(r)=C_0+\alpha r^{-\gamma}$.
We plot the correlation functions $G(r)=\alpha r^{-\gamma}$, i.e., the constants have been subtracted.
Because of the periodic boundary condition, the maximal distance is $r=L/2$.
The calculated exponents $\gamma$ for different $K$ are listed in Table II.
%For $J_2=0$, the from $K=3$ to $K=9$ is 3.15, 2.61, 2.39 and 2.24. For $J_2/J_1=0.5$, $\gamma$ from $K=3$ to $K=9$ is 3.00, 2.53, 2.37 and 2.26.
From these results, one may see that using small filters greatly underestimate the decay length of
the spin correlation functions, and therefore can not represent the ground state very well. Increase filter size may describe the correlation functions more accurately and therefore the better ground state
wave functions and energies. To faithfully represent a quantum state, the size of the convolution filters should cover the correlation length. Otherwise the neural network can not capture the long quantum correlations.

The CQNS is expected to work well for quantum systems with short correlation lengths, e.g., gapped systems. On the other hand, to simulate very large systems, which are much larger than their correlation lengths $\xi$, the CQNS is also advantageous. One may use a suitable filter size $K$, which is larger than $\xi$, but still much less than the system sizes, therefore the number of tunable parameters in CQNS can be greatly reduced.

\section{Conclusions and outlooks}
\label{sec:level5}

We design a convolutional quantum neural network, namely CQNS,
%built from the basic elements of the state-of-the-art convolutional neural networks
to study quantum many-particle problems. By design, CQNS associates
the many-body wave functions with high-order spin correlation functions, which is crucial for its success.
We use CQNS to study the spin-1/2 J$_1$-J$_2$ Heisenberg model which is strongly frustrated and challenging to solve.
We have obtained fairly accurate ground state energies that are even better than previous SBS calculations.\cite{SBS,Replica_exchange} The CQNS is a general and flexible method that can be easily applied to different Hamiltonian, with various boundary conditions (open, periodic and cylindrical, etc.) and in different dimensions. It therefore provides a new powerful tool to study the long-standing quantum many-particle problems.

The CQNS used in this paper is as simple as three layers.
Using deep neural networks may further enhance the state representation ability,
as it is possible to capture the long-range correlation by multiple convolution layers with smaller filters in each layer.
It recently has been proposed that deep CNN may efficiently represent the many-particle states that are even beyond area law.\cite{CNN_TNS} Designing a practical deep CQNS is an interesting and promising route to follow.

\acknowledgments

The authors appreciate insightful discussions with Shao-Jun Dong and Chao Wang. The construction and computation of neural networks in this paper is performed by the mainstream deep learning framework PyTorch. The numerical calculations have been done on the USTC HPC facilities. This work was supported by the National Natural Science Foundation of China (Grant No. 11674306, 61590932), National key R\&D program (No. 2016YFA0301300, 2016YFA0317300).

%
%\section*{References}

%% Here is the endmatter stuff: Supplementary Info, etc.
%% Use \item's to separate, default label is "Acknowledgements"
%\section*{Acknowledgements}

\end{document}